\documentclass{aa}  
\usepackage{graphicx}
\usepackage{txfonts}
\usepackage{xcolor}
\usepackage{hyperref}
\hypersetup{colorlinks, linkcolor=blue, citecolor=blue, urlcolor=blue}
\usepackage{amsmath}
\usepackage{ulem}
\usepackage{soul}
\usepackage{enumitem}

\newcommand{\muvec}{\mbox{\boldmath $\mu$}}

\newcommand{\te}{t_{\rm E}}
\newcommand{\thetae}{\theta_{\rm E}}

\newcommand{\pie}{\pi_{\rm E}}

\newcommand{\dl}{D_{\rm L}}
\newcommand{\ds}{D_{\rm S}}

\definecolor{brown}{rgb}{0.59, 0.29, 0.0}
\definecolor{darkgreen}{rgb}{0.0, 0.42, 0.24}
\definecolor{darkblue}{rgb}{0.01, 0.31, 0.59}
\definecolor{darkblue}{rgb}{0.0, 0.25, 0.42}
\definecolor{blue}{rgb}{0.0,0.0,1.0}
\definecolor{green}{rgb}{0.0,1.0,0.0}

\begin{document}

\title{OGLE-2018-BLG-0584 and KMT-2018-BLG-2119: two microlensing events with
two lens masses and two source stars}
\titlerunning{OGLE-2018-BLG-0584 and KMT-2018-BLG-2119}

\author{
     Cheongho~Han\inst{1} 
\and Andrzej~Udalski\inst{2} 
\and Youn~Kil~Jung\inst{3} 
\and Doeon~Kim\inst{1}
\and Hongjing~Yang\inst{4}     
\\
(Leading authors)\\
     Michael~D.~Albrow\inst{5}   
\and Sun-Ju~Chung\inst{3,6}      
\and Andrew~Gould\inst{7,8}      
\and Kyu-Ha~Hwang\inst{3} 
\and Hyoun-Woo~Kim\inst{3} 
\and Chung-Uk~Lee\inst{3} 
\and Yoon-Hyun~Ryu\inst{3} 
\and Yossi~Shvartzvald\inst{9}   
\and In-Gu~Shin\inst{10} 
\and Jennifer~C.~Yee\inst{10}   
\and Weicheng~Zang\inst{4}     
\and Sang-Mok~Cha\inst{3,11} 
\and Dong-Jin~Kim\inst{3} 
\and Seung-Lee~Kim\inst{3,6} 
\and Dong-Joo~Lee\inst{3} 
\and Yongseok~Lee\inst{3,11} 
\and Byeong-Gon~Park\inst{3,6} 
\and Richard~W.~Pogge\inst{8}
\and Chun-Hwey Kim\inst{12}
\and Woong-Tae Kim\inst{13}
\\
(The KMTNet Collaboration),\\
     Przemek~Mr{\'o}z\inst{2} 
\and Micha{\l}~K.~Szyma{\'n}ski\inst{2}
\and Jan~Skowron\inst{2}
\and Rados{\l}aw~Poleski\inst{2} 
\and Igor~Soszy{\'n}ski\inst{2}
\and Pawe{\l}~Pietrukowicz\inst{2}
\and Szymon~Koz{\l}owski\inst{2} 
\and Krzysztof~A.~Rybicki\inst{2}
\and Patryk~Iwanek\inst{2}
\and Krzysztof~Ulaczyk\inst{14}
\and Marcin~Wrona\inst{2}
\and Mariusz~Gromadzki\inst{2}          
\\
(The OGLE Collaboration)\\
}

\institute{
     Department of Physics, Chungbuk National University, Cheongju 28644, Republic of Korea  \\ \email{cheongho@astroph.chungbuk.ac.kr}     
\and Astronomical Observatory, University of Warsaw, Al.~Ujazdowskie 4, 00-478 Warszawa, Poland                                             
\and Korea Astronomy and Space Science Institute, Daejon 34055, Republic of Korea                                                           
\and Department of Astronomy and Tsinghua Centre for Astrophysics, Tsinghua University, Beijing 100084, China                               
\and University of Canterbury, Department of Physics and Astronomy, Private Bag 4800, Christchurch 8020, New Zealand                        
\and Korea University of Science and Technology, 217 Gajeong-ro, Yuseong-gu, Daejeon, 34113, Republic of Korea                              
\and Max Planck Institute for Astronomy, K\"onigstuhl 17, D-69117 Heidelberg, Germany                                                       
\and Department of Astronomy, The Ohio State University, 140 W. 18th Ave., Columbus, OH 43210, USA                                          
\and Department of Particle Physics and Astrophysics, Weizmann Institute of Science, Rehovot 76100, Israel                                  
\and Center for Astrophysics $|$ Harvard \& Smithsonian 60 Garden St., Cambridge, MA 02138, USA                                             
\and School of Space Research, Kyung Hee University, Yongin, Kyeonggi 17104, Republic of Korea                                              
\and Department of Astronomy \& Space Science, Chungbuk National University, Cheongju 28644, Republic of Korea                              
\and Department of Physics \& Astronomy, Seoul National University, Seoul 08826, Republic of Korea                                          
\and Department of Physics, University of Warwick, Gibbet Hill Road, Coventry, CV4 7AL, UK                                                  
}
\date{Received ; accepted}

\abstract
{}
{
We conduct a systematic investigation of the microlensing data collected during the 
previous observation seasons for the purpose of reanalyzing anomalous lensing events 
with no suggested plausible models.
}
{
We find that two anomalous lensing events OGLE-2018-BLG-0584 and KMT-2018-BLG-2119 cannot be 
explained with the usual models based on either a binary-lens single-source (2L1S) or a 
single-lens binary-source (1L2S) interpretation. We test the feasibility of explaining the 
light curves with more sophisticated models by adding an extra lens (3L1S model) or a source 
(2L2S model) component to the 2L1S lens-system configuration.
}
{
We find that a 2L2S interpretation well explains the light curves of both events, for each 
of which there are a pair of solutions resulting from the close and wide degeneracy. For 
the event OGLE-2018-BLG-0584, the source is a binary composed of two K-type stars, and the
lens is a binary composed of two M dwarfs. For KMT-2018-BLG-2119, the source is a binary
composed of two dwarfs of G and K spectral types, and the lens is a binary composed of a 
low-mass M dwarf and a brown dwarf. 
}
{}


\maketitle

\section{Introduction}\label{sec:one}

Since the pioneering works of the first-generation experiments, for example, OGLE 
\citep{Udalski1994}, MACHO \citep{Alcock1993}, and EROS \citep{Aubourg1993} surveys 
conducted in the early 1990s, searches for light variations of stars induced by 
gravitational lensing have been carried on for more than three decades by multiple 
groups succeeding the early experiments. With the upgrade of instruments and observational 
strategy, the detection rate of lensing events has dramatically increased from a few 
dozens in the early surveys to several thousands in the current lensing surveys that 
are being carried out by the OGLE-IV \citep{Udalski2015}, MOA \citep{Bond2001}, and 
KMTNet \citep{Kim2016} groups.

\begin{table*}[t]
\small
\caption{Microlensing events involved with four or more bodies\label{table:one}}
\begin{tabular}{lllll}
\hline\hline
\multicolumn{1}{c}{Event}                &
\multicolumn{1}{c}{Anomaly type}         &
\multicolumn{1}{c}{Reference}           \\
\hline
OGLE-2006-BLG-109     &  3L1S (multiple planets)     &  \citet{Gaudi2008}, \citet{Bennett2010}        \\    
OGLE-2012-BLG-0026    &  3L1S (multiple planets)     &  \citet{Han2013}, \citet{Beaulieu2016}         \\
OGLE-2018-BLG-1011    &  3L1S (multiple planets)     &  \citet{Han2019}                               \\
OGLE-2019-BLG-0468    &  3L1S (multiple planets)     &  \citet{Han2022f}                              \\
KMT-2021-BLG-1077     &  3L1S (multiple planets)     &  \citet{Han2022a}                              \\
KMT-2021-BLG-0240     &  3L1S (multiple planets)     &  \citet{Han2022d}                              \\
\hline
OGLE-2006-BLG-284     &  3L1S (binary+planet)        &  \citet{Bennett2020}                           \\
OGLE-2007-BLG-349     &  3L1S (binary+planet)        &  \citet{Bennett2016}                           \\
OGLE-2008-BLG-092     &  3L1S (binary+planet)        &  \citet{Poleski2014}                           \\
OGLE-2016-BLG-0613    &  3L1S (binary+planet)        &  \citet{Han2017}                               \\
OGLE-2018-BLG-1700    &  3L1S (binary+planet)        &  \citet{Han2020}                               \\
KMT-2020-BLG-0414     &  3L1S (binary+planet)        &  \citet{Zang2021}                              \\ 
\hline 
KMT-2019-BLG-1715     &  3L2S (binary+planet)        &  \citet{Han2021c}                              \\
\hline
MOA-2010-BLG-117      &  2L2S                        &  \citet{Bennett2018}                           \\
KMT-2018-BLG-1743     &  2L2S                        &  \citet{Han2021a}                              \\
OGLE-2016-BLG-1003    &  2L2S                        &  \citet{Jung2017}                              \\
KMT-2019-BLG-0797     &  2L2S                        &  \citet{Han2021b}                              \\
KMT-2021-BLG-1898     &  2L2S                        &  \citet{Han2022b}                              \\
\hline
\end{tabular}
\tablefoot{ In the case of KMT-2019-BLG-1715, the lens is is composed of three masses and the source is 
a binary.}
\end{table*}

Light curves of most lensing events follow the smooth and symmetric form of a single-lens
single-source (1L1S) event \citep{Paczynski1986}. For a fraction of events, light curves 
exhibit deviations from the 1L1S form, and these deviations are, in most cases, caused by 
the binarity of the lens, 2L1S events \citep{Mao1991}, or the source, 1L2S events 
\citep{Griest1993, Han1997}.

With the increased number of lensing events, it is occasionally found that deviations in 
lensing light curves cannot be explained by the usual 2L1S or 1L2S forms.  One important 
cause of such deviations is the existence of an extra lens component, and thus the lens is 
composed of three masses (3L1S events), as first suggested by \citet{Gaudi1998} and 
theoretically investigated by \citet{Danek2015a, Danek2015b, Danek2019}.  Another major 
cause is that the source is a binary composed of two stars, and thus both the lens and 
source are binaries (2L2S events).  By the time of writing this paper, there exist 18 
lensing events involved with more than four bodies (lens+source), and among them 13 are 
3L1S events, one is 3L2S event, and the other 5 are 2L2S events.  In Table~\ref{table:one}, 
we list these lensing events with brief summary of the anomaly types and related references.

In this paper, we report two lensing events that involve binary lens and binary source 
stars, including OGLE-2018-BLG-0584 and KMT-2018-BLG-2119. In the current lensing surveys,
anomalous lensing events are being analyzed by multiple modelers almost in real time with 
the progress of events, and microlensing models found from these analyses are circulated 
to the microlensing community or posted on web pages.\footnote{For example, the lensing 
event model page maintained by Cheongho Han ({\tt http://astroph.chungbuk.ac.kr/$\sim$cheongho}).}
In this stage, events are mostly analyzed with  the relatively simple 2L1S or 1L2S model, and thus 
events involving four bodies are, in most cases, left without any suggested models describing 
their observed anomalies.  We found the 2L2S nature of OGLE-2018-BLG-0584 and KMT-2018-BLG-2119 
from a systematic investigation of anomalous lensing events for which no plausible 2L1S or 
1L2S models had been previously suggested. Besides these events, we additionally found one event 
(KMT-2021-BLG-1122) produced by a triple-lens system, and the analysis for this 3L1S event 
will be presented in a separate paper.

We present the analyses of the two 2L2S events according to the following organization.  In 
Sect.~\ref{sec:two}, we describe the observations conducted for the individual events and the 
data acquired from the observations. In Sect.~\ref{sec:three}, we explain various lensing models 
tested in the analyses and explain the detailed procedure of the light curve modeling. In the 
subsequent subsections, we present the analyses conducted for the individual events: 
OGLE-2018-BLG-0584 in Sect.~\ref{sec:three-one} and KMT-2018-BLG-2119 in Sect.~\ref{sec:three-two}. 
In Sect.~\ref{sec:four}, we depict the procedures of defining the source stars and estimating the 
angular Einstein radii of the events.  In Sect.~\ref{sec:five}, we mention the Bayesian analyses 
conducted using the observables of the individual lensing events and list the physical lens 
parameters estimated from the analyses.  We summarize results from the analyses and conclude 
in Sect.~\ref{sec:six}.

\section{Observations and data}\label{sec:two}

The two lensing events OGLE-2018-BLG-0584 and KMT-2018-BLG-2119 were found from the surveys 
conducted toward the Galactic bulge in the 2018 season.  Observations of the event OGLE-2018-BLG-0584 
were done by the OGLE and KMTNet groups.  The source of the event lies at $({\rm RA}, 
{\rm DEC})_{\rm J2000}=\textrm{(17:53:36.29, -31:19:42.60)}$, which correspond to the Galactic 
coordinates of $(l, b)=(-1^\circ\hskip-2pt .167, -2^\circ\hskip-2pt .710)$.  The first alert 
of the event was issued by the OGLE group on 2018 April 12 (${\rm HJD}^\prime\equiv 
{\rm HJD}-2450000=8219$), when the source was brighter than the baseline magnitude, 
$I_{\rm base}=19.65$, by $\Delta I\sim 0.3$ mag. The event occurred before the full 
operation of the KMTNet AlertFinder \citep{Kim2018b} system, and it was found from the 
post-season inspection of the data using the KMTNet EventFinder system \citep{Kim2018a}. 
In the KMTNet alert web page\footnote{{\tt https://kmtnet.kasi.re.kr/$\sim$ulens/}}, the event 
is designated as KMT-2018-BLG-2006.  Following the convention of the microlensing community 
using a representative event ID reference of the first discovery group, we hereafter designate the 
event as OGLE-2018-BLG-0584.  The event KMT-2018-BLG-2119, on the other hand, was found 
solely by the KMTNet group from the post-season analysis of the 2018 season data. The 
equatorial and Galactic coordinates of the source are $({\rm RA}, {\rm DEC})_{\rm J2000}=
\textrm{ (17:56:47.62, -28:40:59.99)}$ and $(l, b)= (1^\circ\hskip-2pt .469, -1^\circ\hskip-2pt .975)$, 
respectively. The source is very faint, making it difficult to measure its baseline magnitude 
in the KMTNet template image, but it is registered in the OGLE-III Catalog with a magnitude 
of $I_{\rm cat}=21.38$.

Observations of the events were conducted using the telescopes operated by the OGLE and 
KMTNet lensing surveys. The OGLE group employs a single 1.3~m telescope, and the KMTNet 
group utilizes three identical 1.6~m telescopes for the surveys. The OGLE telescope is 
located at the Las Campanas Observatory in Chile, and the KMTNet telescopes lie at three 
sites of the Siding Spring Observatory in Australia (KMTA), the Cerro Tololo Interamerican 
Observatory in Chile (KMTC), and the South African Astronomical Observatory in South Africa 
(KMTS). The cameras mounted on the OGLE and KMTNet telescopes have 1.4~deg$^2$ and 4~deg$^2$ 
fields of view, respectively.

Images of the source stars were mainly obtained in the $I$ band, and a fraction of images
were acquired in the $V$ band for the source color measurements. Reductions of data and
photometry of the source stars were carried out using the pipelines of the individual groups
developed by \citet{Udalski2003} for the OGLE survey and by \citet{Albrow2009} for the KMTNet
survey.  Additional photometry was done for the KMTC data set using the pyDIA code 
\citep{Albrow2017} to construct color-magnitude diagrams (CMDs) of stars lying around the 
source stars and to estimate the source magnitudes in the $I$ and $V$ passbands.  See more 
detailed discussion in Sect.~\ref{sec:four}.  Following the routine described in \citet{Yee2012}, 
we readjusted the error bars of the photometry data estimated by the pipelines so that the 
error bars are consistent with the scatter of the data and $\chi^2$ per degree of freedom 
(dof) for each data set becomes unity.

\section{Analyses }\label{sec:three}

The analyses of the events were carried out in two steps. In the first step, we modeled the 
light curves of the events with a 2L1S or a 1L2S model. If neither of these models can explain 
the data, we then test more sophisticated models including an extra lens (3L1S model) or source 
(2L2S model) component to the 2L1S lens system configuration.

In the modeling, we search for a lensing solution, which represents a set of lensing parameters 
depicting the configuration of the lens system.  In the simplest case of a 1L1S event, the lensing 
light curve is described by three parameters of $(t_0, u_0, \te)$, which represent the time of the 
closest source approach to the lens, the projected lens-source separation at that time (impact 
parameter), and the Einstein time scale, respectively. The Einstein time scale is defined as 
the time required for a source to transit the angular Einstein radius $\thetae$ of a lens. The 
length of the impact parameter is scaled to $\thetae$.

A 2L1S modeling needs four additional parameters of $(s, q, \alpha, \rho)$ in addition to those of 
the 1L1S model. The first two parameters $s$ and $q$ indicate the projected separation (normalized 
to $\thetae$) and mass ratio between the lens components $M_1$ and $M_2$, respectively, and 
$\alpha$ represents the angle between the direction of the relative lens-source proper motion 
$\muvec$ and the $M_1$--$M_2$ axis (source trajectory angle). The last parameter $\rho$, which 
is defined as the ratio of the angular source radius $\theta_*$ to $\thetae$, that is, $\rho=
\theta_*/\thetae$ (normalized source radius), describes the deformation of a lensing light curve 
by finite-source effects during the crossing of a source over the caustic formed by a binary 
lens system \citep{Bennett1996}.

A 1L2S modeling also requires one to include extra parameters.  These extra parameters are 
$(t_{0,2}, u_{0,2}, q_F)$, in which the first two represent the peak time and impact parameter 
of the source companion ($S_2$) to the primary source ($S_1$), and the last parameter denotes 
the flux ratio between $S_1$ and $S_2$ \citep{Hwang2013}. 
In the 1L2S modeling, we use the notations $(t_{0,1}, 
u_{0,1})$ to designate the parameters related to $S_1$.

Adding a tertiary lens component ($M_3$) to a 2L1S configuration requires one to include three 
extra parameters of $(s_3, q_3, \psi)$ \citep{Han2013}.  These parameters represent the projected 
separation and mass ratio between $M_1$ and $M_3$, and the position angle of the third body as 
measured from the $M_1$--$M_2$ axis centered at the position of $M_1$, respectively. In order 
to distinguish the parameters related to $M_2$ from those describing $M_3$, we use the notations 
$(s_2, q_2)$ to designate the separation and mass ratio between $M_1$ and $M_2$.

\begin{table}[t]
\small
\caption{Lensing parameters of tested models\label{table:two}}
\begin{tabular*}{\columnwidth}{@{\extracolsep{\fill}}llr}
\hline\hline
\multicolumn{1}{c}{Model}             &
\multicolumn{1}{c}{Parameters }       &
\multicolumn{1}{c}{$N_{\rm par}$}     \\
\hline
 1L1S  &  $t_0$, $u_0$, $\te$                                                                                 &   3 \\
 2L1S  &  $t_0$, $u_0$, $\te$, $s$, $q$, $\alpha$, $\rho$                                                     &   7 \\
 1L2S  &  $t_{0,1}$, $u_{0,1}$, $t_{0,2}$, $u_{0,2}$, $\te$, $q_F$                                            &   6 \\
 3L1S  &  $t_{0}$, $u_0$, $\te$, $s_2$, $q_2$, $\alpha$, $s_3$, $q_3$, $\psi$, $\rho$                         &  10 \\
 2L2S  &  $t_{0,1}$, $u_{0,1}$, $t_{0,2}$, $u_{0,2}$, $\te$, $s$, $q$, $\alpha$, $\rho_1$, $\rho_2$, $q_F$    &  11 \\
\hline
\end{tabular*}
\end{table}

Similarly, adding a second source to a 2L1S configuration also requires extra parameters in 
modeling. These parameters are $(t_{0,2}, u_{0,2}, \rho_2, q_F)$. Here we use the subscript 
"2" to designate the parameters related to the second source, and the subscript "1" to 
denote the parameters related to the primary source, that is, $(t_{0,1}, u_{0,1}, \rho_1)$.  
In Table~\ref{table:two}, we summarize the lensing parameters of the models tested in our 
analyses together with the total numbers of parameters, $N_{\rm par}$, included in the individual 
models.

The three-body (2L1S and 1L2S) modeling was conducted considering the patterns of the anomalies
appearing in the light curves. In the 2L1S modeling, we initially searched for the binary
parameters $(s, q)$ via a grid approach, while the other parameters were found via a downhill
approach using a Markov Chain Monte Carlo (MCMC) logic, constructed a $\chi^2$ map on the
$\log s$--$\log q$ plane, identified local solutions on the $\chi^2$ map, and then refined the 
individual local solutions by allowing all parameters to vary. In the 1L2S modeling, we first 
fit the light curve with a 1L1S model, obtained approximate values of the 1L1S parameters, and 
then tested various configurations of the second source considering the time, magnitude, and 
pattern of the anomaly.  As will be discussed in the following subsections, it is found that 
the light curves of the two events OGLE-2018-BLG-0584 and KMT-2018-BLG-2119 cannot be precisely 
described by either of the three-body models.

Although the anomalies cannot be fully described by three-body models, we find that 2L1S models
can partially describe the anomalies for both events. The light curve of a 2L2S event is the
superposition of the light curves of the two 2L1S events involved with the individual source 
stars.  Similarly, the anomalies induced by a triple-lens system, in many cases, are known to 
be approximated as the superposition of the anomalies induced by the two binary pairs, that is,
$M_1$--$M_2$ and $M_1$--$M_3$ pairs \citep{Bozza1999, Han2001}. The four-body modeling was conducted 
under this superposition approximation by first finding a 2L1S model of each event describing a
part of the anomaly. Based on this 2L1S model, we conducted a 2L2S modeling by testing various
trajectories of the second source considering the anomaly part that could not be explained by 
the 2L1S model. In the 3L1S model, we first conducted grid searches for the perimeters related 
to $M_3$, that is, $(s_3, q_3, \psi)$, with the other parameters fixed as the values found from 
the 2L1S model, and then refined the lensing solutions found from the grid search by letting all 
parameters vary.  In the following subsections, we explain details of the modeling conducted for 
the two events and present the best solutions explaining the anomalies of the individual events.

\begin{figure}[t]
\includegraphics[width=\columnwidth]{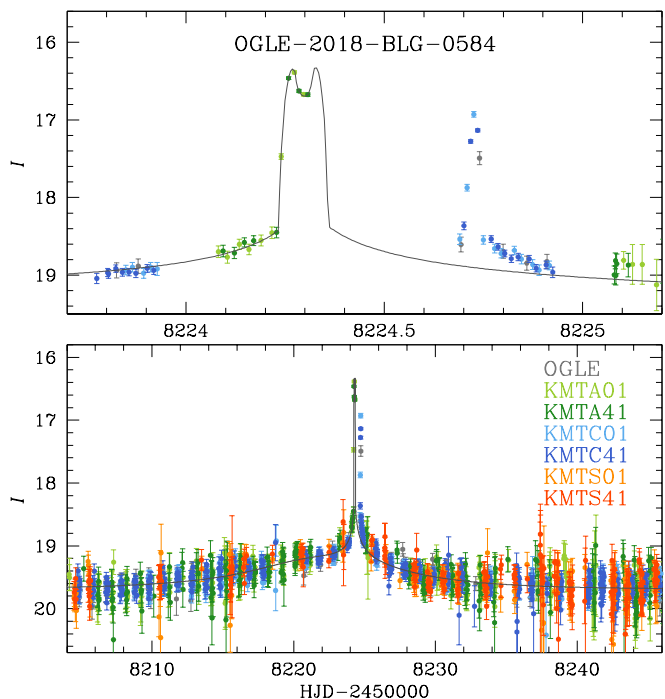}
\caption{
Light curve of OGLE-2018-BLG-0584.  The curve drawn over the data points is the model of the 
close 2L1S solution found from the fit of the data excluding those in the region $8224.5 \leq 
{\rm HJD}-2450000 \leq 8225.5$. The upper panel shows the zoom of the anomaly region.
}
\label{fig:one}
\end{figure}

\subsection{OGLE-2018-BLG-0584}\label{sec:three-one}

The light curve of the lensing event OGLE-2018-BLG-0584 is shown in Figure~\ref{fig:one}. It 
is characterized by two distinctive anomaly features centered at ${\rm HJD}^\prime \sim 8224.3$ 
($t_1$) and $\sim 8224.7$ ($t_2$). From the sharp rise and fall of the source flux together with 
the non-smooth curvature in the light curve around the anomalies, it is likely that both anomalies 
are produced by caustic crossings of a source.  The event lies in the two overlapping prime KMTNet 
fields of BLG01 and BLG41, toward which observations were conducted with 
a combined cadence of 0.25~hr, and thus the rising part of the first anomaly feature and both the 
rising and falling parts of the second anomaly feature were densely covered by the data.

\begin{figure}[t]
\includegraphics[width=\columnwidth]{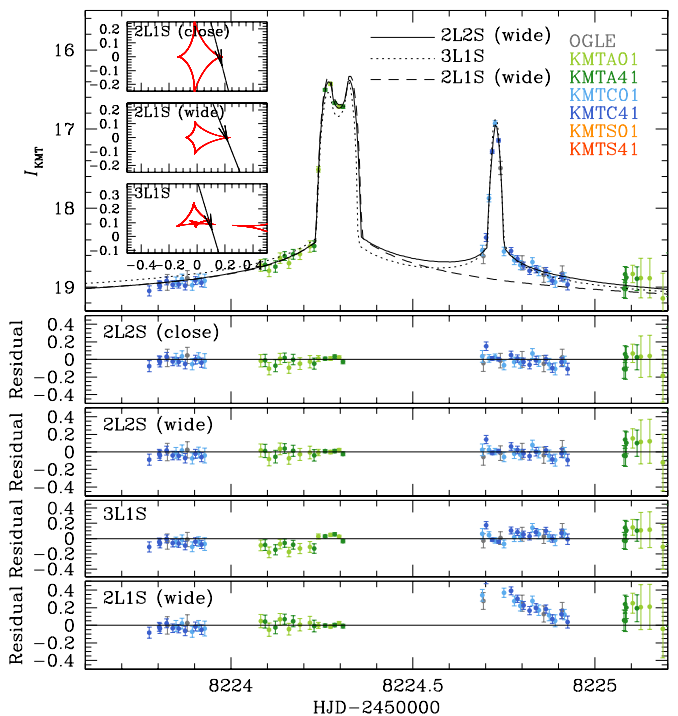}
\caption{
Comparisons of models in the anomaly region of the OGLE-2018-BLG-0584 light curve.  The top 
panel shows the four models of the close and wide 2L2S, 3L1S, and 2L1S solutions, and the lower 
panels show the residuals from the individual models.  The three insets in the top panel show 
the lens system configurations of the 2L1S and 3L1S solutions. In each inset, the red figures 
represent the caustics and the line with an arrow indicates the source trajectory.  The 
configuration of the 2L2S solutions are presented in Fig.~\ref{fig:three}. 
}
\label{fig:two}
\end{figure}

Because caustics are produced by a multiple lens system, we exclude the 1L2S interpretation and 
start the analysis of the light curve with a 2L1S model.  Despite a thorough investigation of the 
parameter space, we found no 2L1S solution that simultaneously described both anomalies. We then 
checked whether a 2L1S model could describe either of the anomalies. For this check, we conducted 
an additional 2L1S modeling by fitting the light curve with the exclusion of the data around the 
second anomaly lying in the range of $8224.5 \leq {\rm HJD}^\prime \leq 8225.5$. From this, we 
found a pair of 2L1S models that could describe the first anomaly of the light curve. The two 
solutions with binary parameters of $(s, q)_{\rm close}\sim (0.6, 0.6)$ and $(s, q)_{\rm wide}
\sim (2.4, 1.1)$ result from the close--wide degeneracy \citep{Dominik1999, An2005}.  In 
Figures~\ref{fig:one} and \ref{fig:two}, we draw the model curve over the data points for one 
(wide solution) of the two 2L1S solutions. The lens-system configurations of the two 2L1S 
solutions are shown in the insets of the top panel in Figure~\ref{fig:two}.

\begin{table}[t]
\small
\caption{2L2S models of OGLE-2018-BLG-0584\label{table:three}}
\begin{tabular*}{\columnwidth}{@{\extracolsep{\fill}}lrr}
\hline\hline
\multicolumn{1}{c}{Parameter}   &
\multicolumn{1}{c}{Close}       &
\multicolumn{1}{c}{Wide}       \\
\hline
 $\chi^2/{\rm dof}$         &  $8954.6/8957          $   & $8952.3/8957        $            \\
 $t_{0,1}$ (HJD$^\prime$)   &  $8223.851 \pm 0.026   $   & $8222.996 \pm 0.017 $            \\
 $u_{0,1}$                  &  $0.167 \pm 0.003      $   & $0.217 \pm 0.001    $                             \\
 $t_{0,2}$ (HJD$^\prime$)   &  $8224.254 \pm 0.027   $   & $8223.310 \pm 0.014 $            \\
 $u_{0,2}$                  &  $0.177 \pm 0.003      $   & $0.236 \pm 0.001    $         \\
 $\te$ (days)               &  $9.30 \pm 0.11        $   & $14.78 \pm 0.13     $      ($10.23 \pm 0.090 $)   \\
 $s$                        &  $0.640 \pm 0.006      $   & $2.370 \pm 0.002    $            \\
 $q$                        &  $0.600 \pm 0.020      $   & $1.085 \pm 0.059    $            \\
 $\alpha$ (rad)             &  $4.432 \pm 0.012      $   & $4.327 \pm 0.004    $            \\
 $\rho_1$ ($10^{-3}$)       &  $1.88 \pm 0.06        $   & $1.28 \pm 0.05      $      ($1.84 \pm 0.07 $)   \\
 $\rho_2$ ($10^{-3}$)       &  $1.30 \pm 0.11        $   & $0.93 \pm 0.07      $      ($1.34 \pm 0.10 $)   \\
 $q_F$                      &  $0.29 \pm 0.01        $   & $0.27 \pm 0.01      $         \\
\hline
\end{tabular*}
\tablefoot{ 
The values of $\te$, $\rho_1$ and $\rho_2$ of the wide solution presented in the parentheses 
are scaled to the Einstein radius of the lens component lying closer to the source trajectory.
}
\end{table}

We further checked whether the data around the second anomaly could be explained with the 
introduction of a tertiary lens component.  The model curve (dotted curve) and residual of the 
best-fit 3L1S solution are presented in Figure~\ref{fig:two}. We also present the lens-system 
configuration of the 3L1S solution in the inset of the top panel. It is found that the 3L1S model 
approximately describes both anomaly features at around $t_1$ and $t_2$, but  it leaves 
systematic subtle negative residuals in the region around the first anomaly and positive residuals 
in the region around the second anomaly, indicating that another interpretation is needed for the 
precise description of the anomalies.

\begin{figure}[t]
\includegraphics[width=\columnwidth]{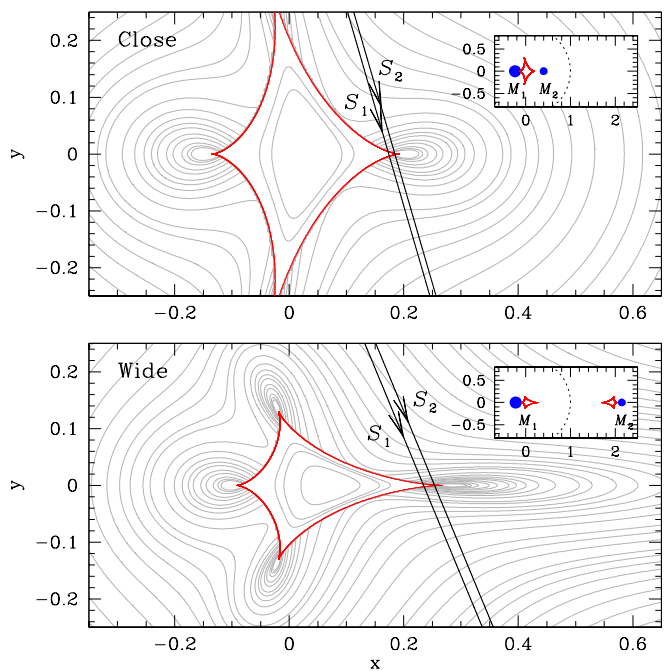}
\caption{
Lens-system configurations for the close (upper panel) and wide (lower panel) 2L2S solutions of 
OGLE-2018-BLG-0584.  In each panel, The close figure represent the caustic, and the two lines 
with arrows marked by $S_1$ and $S_2$ represent the trajectories of the primary ($S_1$) and 
secondary ($S_2$) source stars, respectively.  The grey curves encompassing the caustic 
represent the equi-magnification contours.  The inset in each panel shows the whole view of 
the lens system, where the blue dots represent the positions of the binary lens components
and the dotted circle is the Einstein ring.
}
\label{fig:three}
\end{figure}

We additionally tested a 2L2S interpretation of the anomalies by adding an extra source component 
to the 2L1S model.  From this, it was found that both anomalies were well explained by a 2L2S 
interpretation. We identified two 2L2S solutions, which resulted from the initial parameters of 
the close and wide 2L1S solutions. The lensing parameters of the individual solutions, which we 
refer to as "close" ($s>1.0$) and "wide" ($s>1.0$) solutions, are listed in Table~\ref{table:three} 
together with the values of $\chi^2/{\rm dof}$.  For the values of $\te$, $\rho_1$, and $\rho_2$ 
of the wide solution, we additionally present the values scaled to the Einstein radius of the lens 
component lying closer to the source trajectory, values in the parentheses, to show that these 
values are similar to those of the close solution.  The fits of the two solutions are nearly the 
same, and the wide solution is preferred over the close solution by merely $\Delta\chi^2=2.3$. 
Although the binary parameters $(s, q)$ of the two degenerate solutions are substantially 
different from each other, the flux ratios between the source stars, $q_F\sim 0.28$, estimated 
by the two degenerate solutions are similar to each other.

The lens-system configuration of the 2L2S model is shown in Figure~\ref{fig:three}. The two lines 
with arrows represent the source trajectories of the primary (marked by "$S_1$") and secondary ("$S_2$") 
source stars.  The configuration is similar to those of the corresponding 2L1S models, shown in the 
top-panel insets of Figure~\ref{fig:two}, except that there is an additional trajectory of the second 
source, which trails the primary source with a small separation.  According to the models, the anomaly 
at $t_2$, which could not be explained by the 2L1S models, was produced by the crossing of $S_2$ over 
the tip of the caustic with an impact parameter slightly greater than that of $S_1$. It is found that 
the 2L2S model yields a better fit than the 3L1S model by $\Delta\chi^2=220.1$, and this strongly 
supports the 2L2S interpretation of the anomaly.

According to the close 2L2S solution, the separation between the two source stars is
$\Delta u = \{ (u_{0,1}-u_{0,2})^2 + [(t_{0,1}-t_{0,2})/\te]^2 \}^{1/2}\sim 0.044$.
As will be discussed in Sect.~\ref{sec:four}, the angular Einstein radius is $\thetae\sim 0.57$~mas.  
By adopting the distance to the source of $\ds\sim 8$~kpc, then the two stars of the binary source are 
separated by $a_\perp = \Delta u \ds\thetae\sim  0.20$~AU in projection, and probably $a\sim 0.25$~AU 
in 3-dimensional space.  In this case, the orbital period of the source would be about $P = 
\sqrt{(a^3/M_{\rm S, tot})} \sim 35$~days, where we adopt the total mass of the source of $M_{\rm S, 
tot}\sim 1.8~M_\odot$ considering the stellar types of the source stars to be discussed in 
Sect.~\ref{sec:four}.  This orbital period is long enough to ignore source orbital motion because 
anomaly lasted only about 2 days.

\begin{figure}[t]
\includegraphics[width=\columnwidth]{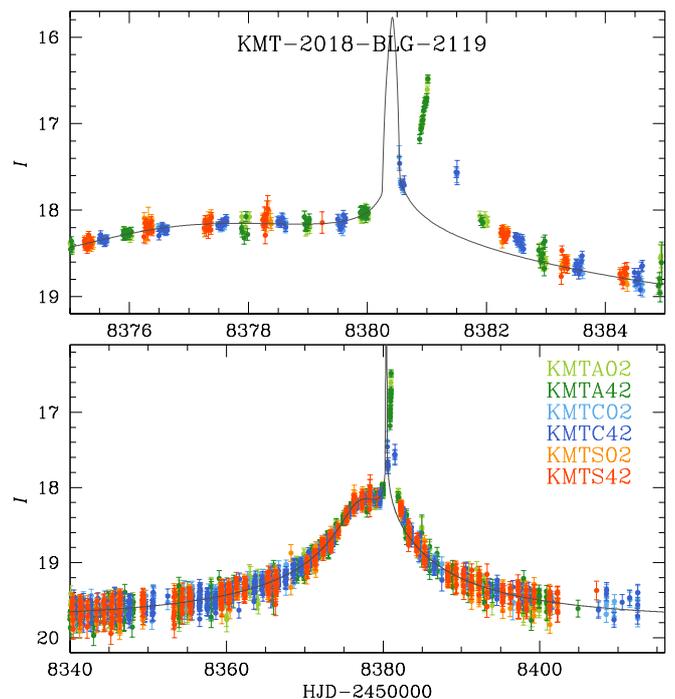}
\caption{
Light curve of KMT-2018-BLG-2119.  The curve drawn over the data points is a 2L1S
model (close solution) obtained by fitting the light curve with the exclusion of the data in the
region of $8380.7\leq {\rm HJD}^\prime \leq 8384.0$. 
}
\label{fig:four}
\end{figure}

\subsection{KMT-2018-BLG-2119}\label{sec:three-two}

The light curve of the lensing event KMT-2018-BLG-2119 is presented in Figure~\ref{fig:four}. It
shows a strong anomaly appearing about two days after the peak, and the anomaly is characterized 
by three features: a weak bump at ${\rm HJD}^\prime \sim 8378$ ($t_1$) and two strong features at 
$\sim 8380.4$ ($t_2$) and $\sim 8381.0$ ($t_3$). From the discontinuous derivatives of the source 
flux, the features around $t_2$ and $t_3$ are likely to be involved with a caustic, and this 
excludes the 1L2S interpretation of the light curve. Despite the fact that the source of the event 
lies in the two overlapping KMTNet prime fields of BLG02 and BLG42, toward which the event was 
covered with a combined cadence of 0.25~hr, the features at around the peak of the anomaly were 
only partially covered. This was not only because the observing window at around the time of the 
anomaly (around September 17) was short but also because the sky in South Africa was clouded out 
during the anomaly.  

\begin{figure}[t]
\includegraphics[width=\columnwidth]{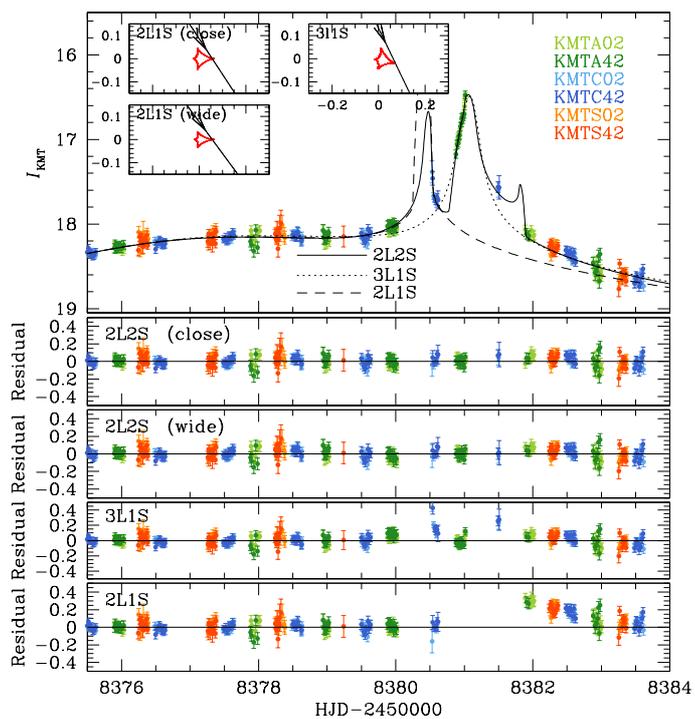}
\caption{
Comparison of models in the anomaly
region of the KMT-2018-BLG-2119 light curve. Notations are the same as those in Fig.~\ref{fig:two}. 
}
\label{fig:five}
\end{figure}

\begin{table}[t]
\small
\caption{2L2S models of KMT-2018-BLG-2119\label{table:four}}
\begin{tabular*}{\columnwidth}{@{\extracolsep{\fill}}lrr}
\hline\hline
\multicolumn{1}{c}{Parameter}   &
\multicolumn{1}{c}{Close}       &
\multicolumn{1}{c}{Wide}       \\
\hline
 $\chi^2/{\rm dof}$         &  $7573.1/7620        $  &  $7587.6/7620        $  \\
 $t_{0,1}$ (HJD$^\prime$)   &  $8378.760 \pm 0.037 $  &  $8378.550 \pm 0.035 $  \\
 $u_{0,1}$                  &  $0.053 \pm 0.002    $  &  $0.048 \pm 0.002    $  \\
 $t_{0,2}$ (HJD$^\prime$)   &  $8380.982 \pm 0.008 $  &  $8380.978 \pm 0.013 $  \\
 $u_{0,2}$                  &  $-0.012 \pm 0.001   $  &  $-0.010 \pm 0.001   $  \\
 $\te$ (days)               &  $39.03 \pm 1.24     $  &  $45.56 \pm 1.37     $  \\
 $s$                        &  $0.547 \pm 0.008    $  &  $2.113 \pm 0.043    $  \\
 $q$                        &  $0.058 \pm 0.003    $  &  $0.063 \pm 0.004    $  \\
 $\alpha$ (rad)             &  $3.861 \pm 0.009    $  &  $3.850 \pm 0.008    $  \\
 $\rho_1$ ($10^{-3}$)       &  $1.70 \pm 0.40      $  &  $1.88 \pm 0.44      $  \\
 $\rho_2$ ($10^{-3}$)       &  $1.26 \pm 0.15      $  &  $1.06 \pm 0.10      $  \\
 $q_{F,I}$                  &  $0.20 \pm 0.01      $  &  $0.22 \pm 0.01      $  \\
\hline
\end{tabular*}
\end{table}

As in the case of OGLE-2018-BLG-0584, it was found that a 2L1S model could not precisely describe 
all the anomaly features.  In order to check whether a 2L1S model can partially describe the 
anomaly, we divided the anomaly into two parts, in which the first part includes the features 
around $t_1$ and $t_2$, and the other part includes the feature around $t_3$, and then fit the 
light curve by excluding the data around the second part lying in the range of $8380.7\leq 
{\rm HJD}^\prime \leq 8384.0$. From this modeling, we found that the anomaly features in the first 
part were well explained by a pair of 2L1S models resulting from the close--wide degeneracy with 
$(s, q)_{\rm close}\sim (0.32, 0.13)$ and $(s, q)_{\rm wide}\sim (2.53, 0.11)$.  The model curve 
of the close solution is drawn over the data points in Figures~\ref{fig:four} and \ref{fig:five}, and 
the lens-system configurations of the close and wide solutions are presented in the insets of the 
top panel in Figure~\ref{fig:five}. According to these solutions, the first part of the anomaly is 
explained by the source approach close to the upper cusp of the caustic, producing the weak anomaly 
at around $t_1$, and the passage through the protruding right-side tip of the caustic, producing the 
sharp anomaly feature at around $t_2$.  The KMTC data at around $t_2$ correspond to the falling side 
of the caustic crossing.

For the explanation of the whole anomaly features, we then checked a 3L1S interpretation by 
modeling the light curve with the use of the initial lensing parameters as those found from 
the 2L1S modeling.  The model curve of the best-fit 3L1S solution and its residual are shown 
in Figure~\ref{fig:five} together with the lens-system configuration of the solution, shown in 
the inset of the top panel.  It was found that the model could not precisely describe the anomaly 
features, although it approximately delineated the feature at around $t_3$.

\begin{table*}[t]
\small
\caption{Source colors, magnitudes, angular radii, Einstein radii, and relative proper motion \label{table:five}}
\begin{tabular}{lllll}
\hline\hline
\multicolumn{1}{c}{Quantity}            &
\multicolumn{2}{c}{OGLE-2018-BLG-0584}  &
\multicolumn{2}{c}{KMT-2018-BLG-2119}   \\
\multicolumn{1}{c}{}                    &
\multicolumn{1}{c}{$S_1$}               &
\multicolumn{1}{c}{$S_2$}               &
\multicolumn{1}{c}{$S_1$}               &
\multicolumn{1}{c}{$S_2$}               \\
\hline
 $(V-I)_S$               &  $1.949 \pm 0.164  $   &    $2.483 \pm 0.564   $   &  $2.176 \pm 0.093  $       &   $2.385 \pm 0.480  $     \\    
 $I_S$                   &  $20.659\pm 0.037  $   &    $22.175 \pm 0.044  $   &  $21.474 \pm 0.014 $       &   $23.113 \pm 0.049 $     \\    
 $(V-I, I)_{\rm RGC}$    &  $(2.239, 16.506)  $   &    $\leftarrow        $   &  $(2.560, 16.272)  $       &   $\leftarrow       $     \\
 $(V-I, I)_{\rm RGC,0}$  &  $(1.060, 1.484)   $   &    $\leftarrow        $   &  $(1.060, 14.365)  $       &   $\leftarrow       $     \\
 $(V-I)_{S,0}$           &  $0.770 \pm 0.164  $   &    $1.304 \pm 0.564   $   &  $0.676 \pm 0.093  $       &   $0.885 \pm 0.476  $     \\
 $I_{S,0}$               &  $18.637 \pm 0.037 $   &    $20.153 \pm 0.044  $   &  $19.567 \pm 0.014)$       &   $21.206 \pm 0.049 $     \\
\hline          
 $\theta_*$ ($\mu$as)    &  $0.63 \pm 0.11    $   &    $0.55 \pm 0.31     $   &  $0.37 \pm 0.04    $       &   $0.22 \pm 0.11     $    \\
 $\thetae$ (mas)         &  $0.48 \pm 0.09    $   &    $0.60 \pm 0.34     $   &  $0.17 \pm 0.04    $       &   $0.15 \pm 0.07     $    \\
 $\mu$ (mas/yr)          &  $12.18 \pm 2.21   $   &    $15.10 \pm 8.64    $   &  $1.74 \pm 0.40    $       &   $1.57 \pm 0.77     $    \\
\hline
\end{tabular}
\end{table*}

We further tested a 2L2S interpretation of the anomaly by adding an extra source to the 2L1S
system. From this, it was found that all the features of the anomaly were well explained by a 
2L2S model. We found a pair of solutions obtained with the initial parameters of the close and 
wide 2L1S solutions. The full lensing parameters of the "close" and "wide" solutions are listed in 
Table~\ref{table:four}, and the model curve of the close solution and residuals of both solutions 
are shown in Figure~\ref{fig:five}. The close solution yields a modestly better fit to the 
data than the wide solution: $\Delta\chi^2 =14.5$.  The binary-lens parameters are 
$(s, q)_{\rm close}\sim (0.55, 0.06)$ and $(s, q)_{\rm wide}\sim (2.11, 0.06)$ for the close 
and wide solutions, respectively. Considering that typical lensing events detected toward the 
Galactic bulge fields are generated by low-mass stars \citep{Han2003}, the low mass ratio 
between the lens components suggests that the lens companion is likely to be a brown dwarf.  
The flux from the second source corresponds to about 20\% of the flux from the primary source, 
that is, $q_F\sim 0.2$.  The orbital period of the binary source, estimated in a similar 
fashion to that of the OGLE-2018-BLG-0584 binary source is $P\sim 14$~days.  Considering that 
the anomaly features are separated by $\sim 1$~day, the orbital motion of the source does not 
significantly affect the anomaly.  The effect of the lens orbital motion would be even smaller
because the orbital period of the lens is $P\sim 2$~yr, which is based on the lens mass and binary 
separation estimated in Sect.~\ref{sec:five}, even for the close solution.

\begin{figure}[t]
\includegraphics[width=\columnwidth]{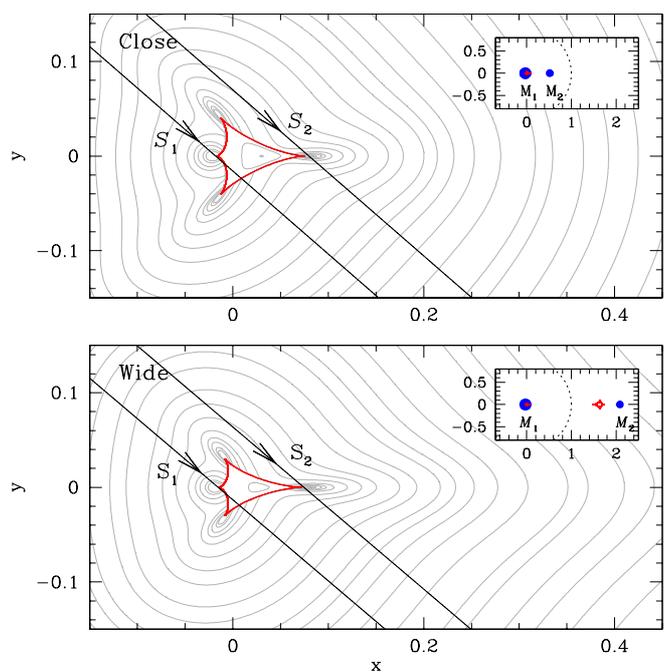}
\caption{
Lens-system configurations for the close and wide 2L2S solutions of KMT-2018-BLG-2119.
Notations are the same as those in Fig.~\ref{fig:three}. 
}
\label{fig:six}
\end{figure}

Figure~\ref{fig:six} shows the lens-system configurations of the close (upper panel) and wide 
(lower panel) 2L2S solutions. The caustics of the individual solutions are similar to those of 
the corresponding 2L1S solutions presented in the insets of Figure~\ref{fig:five}. For both close 
and wide 2L2S solutions, the features in the second half of the anomaly are explained by an extra 
source. The second source approached the left on-axis cusp of the caustic, and then successively 
crossed the lower left and right folds of the caustic.  The anomaly feature at around $t_3$,
covered by the KMTA data set, was produced from the combination of the cusp approach and caustic 
entrance of the second source, and the three KMTC data points at ${\rm HJD}^\prime \sim 8381.5$ 
correspond to the U-shape region between the caustic entrance and exit of the second source, 
although the caustic exit was not covered by the data.

\section{Source stars and Einstein radii}\label{sec:four}

In this section, we specify the source stars of the events for the estimation of the angular 
Einstein radii as well as for the full characterization of the events. We specify the source 
of each event by measuring the extinction- and reddening-corrected (dereddened) color and magnitude 
using the \citet{Yoo2004} routine. In this routine, the source location in the instrumental CMD of 
neighboring stars around the source is first determined by measuring the instrumental magnitudes 
of the source in two passbands, $I_S$ and $V_S$, and then the source color and magnitude, 
$(V-I, I)_S$, are calibrated using the centroid of red giant clump (RGC), with the instrumental 
color and magnitude of $(V-I, I)_{\rm RGC}$, in the CMD.  The RGC centroid is used as a 
reference for calibration because its dereddened color and magnitude, $(V-I, I)_{\rm RGC,0}$, 
are known \citep{Bensby2013, Nataf2013}.

\begin{figure}[t]
\includegraphics[width=\columnwidth]{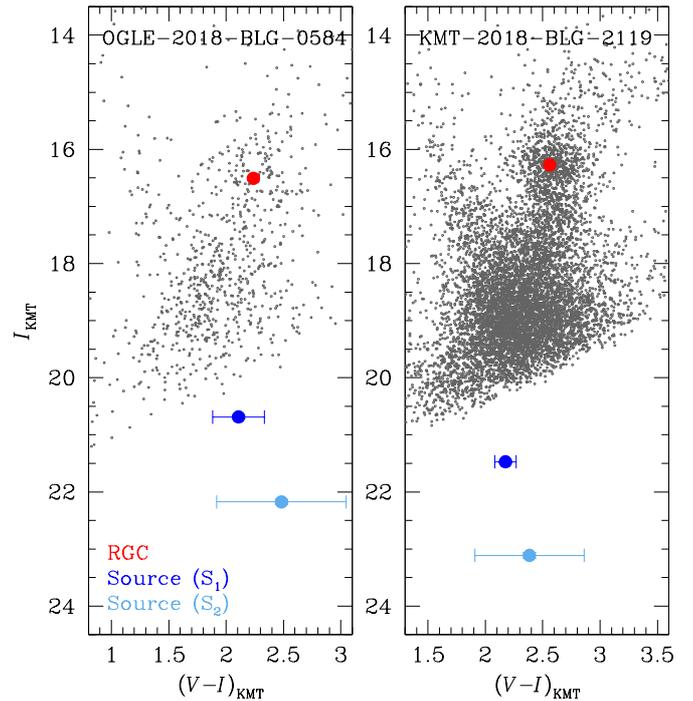}
\caption{
Locations of the binary source stars ($S_1$ and $S_2$) in the instrumental color-magnitude 
diagrams of neighboring stars with respect to the centroids of red giant clump (RGC) for 
the lensing events OGLE-2018-BLG-0584 (left panel) and KMT-2018-BLG-2119 (right panel). 
}
\label{fig:seven}
\end{figure}

Figure~\ref{fig:seven} shows the source locations of the two events in the instrumental CMDs 
constructed with the KMTC data sets processed using the pyDIA code.  For each event, we first 
measured the combined magnitudes of the source $I_S$ and $V_S$, with the flux values of $F_{S,I}$ 
and $F_{S,V}$, respectively, by regressing the pyDIA light curve data measured in the individual 
passbands with respect to the flux predicted by the model of 
\begin{equation}
F_{p}(t) = F_{S,p}  [A_1(t) + q_{F,p} A_2(t)] + F_{b,p},
\label{eq1}
\end{equation}
where $A_1(t)$ and $A_2(t)$ represent the lensing magnifications involved with the primary 
and secondary source stars, respectively, $F_{S,p}$ is the combined flux from the two source 
stars, $F_{b,p}$ represents the blended flux from nearby unresolved stars, and the subscript 
"$p$" denotes the observation passband, that is, $I$ and $V$.  With the measured $F_{S,p}$, 
we then estimated the flux values of the individual source components $S_1$ and $S_2$ using 
the relations 
\begin{equation}
F_{S_1,p}=\left( {1\over 1+q_{F,p}}\right)F_{S,p};\qquad
F_{S_2,p}=\left( {q_{F,p}\over 1+q_{F,p}}\right)F_{S,p}, 
\label{eq2}
\end{equation}
where $F_{S,p}=F_{S_1,p}+F_{S_2,p}$.
The $V$-band flux ratios, which are
$q_{F,V}=0.18\pm 0.15$ for KMT-2018-BLG-0584 and
$q_{F,V}=0.18\pm 0.09$ for KMT-2018-BLG-2119, 
were measured from the additional modeling including the $V$-band data of the 
individual events. 
The instrumental color and magnitude were then calibrated 
by $(V-I, I)_0 = (V-I, I)_{\rm RGC,0} + \Delta (V-I, I)$, where $\Delta (V-I, I) = (V-I, I)_S - 
(V-I)_{\rm RGC}$ denotes the color and magnitude offsets of the source from the RGC centroid.

In Table~\ref{table:five}, we list the values of $(V-I, I)_S$, $(V-I, I)_{\rm RGC}$, 
$(V-I, I)_{\rm RGC,0}$, and $(V-I, I)_{S,0}$ of the stars comprising the binary sources of 
the two events.  According to the estimated colors and magnitudes, the source of OGLE-2018-BLG-0584 
is a binary composed of two K-type dwarfs, and that of KMT-2018-BLG-2119 is a binary composed of 
two dwarfs of G and K spectral types.

With the estimated source color and magnitude, the angular Einstein radius was estimated from 
the relation 
\begin{equation}
\thetae={\theta_*\over \rho},
\label{eq3}
\end{equation}
where the angular radius of the source star, $\theta_*$, was deduced from the de-reddened color 
and magnitude, and the normalized source radius is obtained from modeling. For this, we converted 
the $V-I$ color into the $V-K$ color using the \citet{Bessell1988} relation, and then estimated 
$\theta_*$ from the \citet{Kervella2004} relation between $(V-K, V)$ and $\theta_*$. With the 
measured Einstein radius, the relative lens-source motion was estimated from $\thetae$ and $\te$ by
\begin{equation}
\mu={\thetae\over \te}.
\label{eq4}
\end{equation}

The estimated values of $\theta_*$, $\thetae$ and $\mu$ are presented in Table~\ref{table:five}. In the 
table, we list two sets of $(\theta_*, \thetae, \mu)$ values, in which one set is estimated based on the 
color and magnitude of the primary source $S_1$ (the values presented in the column with the heading 
"$S_1$") and the other set is estimated based on those of the secondary source $S_2$ (in the column 
with the heading "$S_2$").  It is found that the values $\thetae$ and $\mu$ estimated from $S_1$ 
and $S_2$ are consistent, and this gives more credibility to the 2L2S interpretations of the events.

We note that the estimated lens-source proper motion of  OGLE-2018-BLG-0584, $\mu> 12.2$~mas/yr,
is substantially bigger than $\sim 5$~mas/yr of typical lensing events.  The lens of this event can 
be resolved from the source in about 10 years by high angular-resolution observations with 8~m class 
telescopes or the Hubble Space Telescope, as in the case of the planetary event OGLE-2005-BLG-169 
\citep{Batista2015, Bennett2015}.  The lens luminosity measurement from this high resolution image 
will be useful not only to confirm the 2L2S interpretation but also to constrain the physical lens 
parameters.  For the potential follow-up observations in the future, we estimate that the $K$-band 
magnitude of the combined source stars would be $K\sim 17.5$.

\begin{figure}[t]
\includegraphics[width=\columnwidth]{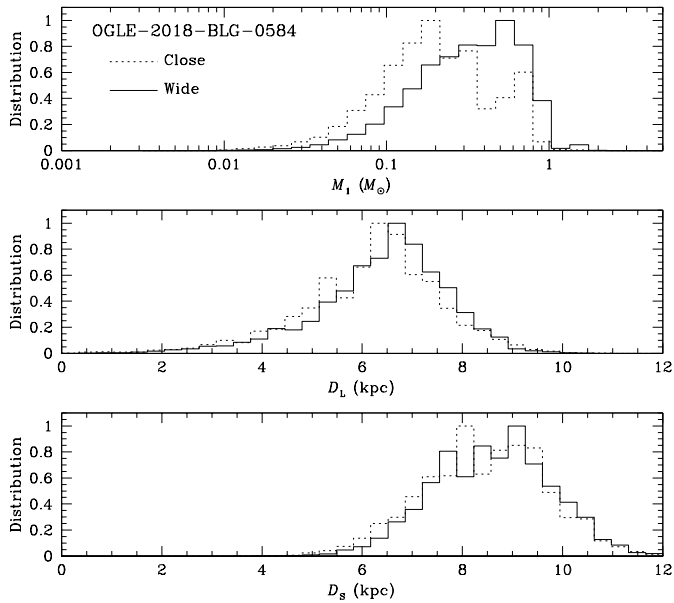}
\caption{
Bayesian posteriors of the primary lens
mass, distance to the lens and source for the lensing
event OGLE-2018-BLG-0584. Dotted and solid curves are based on the close and wide solutions,
respectively. 
}
\label{fig:eight}
\end{figure}

\section{Physical lens parameters}\label{sec:five}

In this section, we estimate the physical parameters of the lens systems including the mass and
distance. These parameters can be uniquely determined by measuring the extra lensing observables 
of the microlens parallax $\pie$ and Einstein radius by 
\begin{equation}
M={\thetae\over \kappa \pie};\qquad
\dl = {{\rm AU}\over \pie\thetae + \pi_{\rm S}},
\label{eq5}
\end{equation}
where $\kappa =4G/(c^2{\rm AU}) \simeq 8.14~{\rm mas}/M_\odot$, $\pi_{\rm S} ={\rm AU}/D_{\rm S}$, 
and $D_{\rm S}$ denotes the distance to the source. For both events OGLE-2018-BLG-0584 and 
KMT-2018-BLG-2119, the Einstein radii were measured, but the values of the microlens parallax could 
not be securely measured for either of the events. We, therefore, estimate $M$ and $\dl$ by conducting Bayesian 
analyses based on the measured observables of $\te$ and $\thetae$, which are related to the physical 
parameters by
\begin{equation}
\te = {\thetae\over \mu};\qquad
\thetae =(\kappa M \pi_{\rm rel})^{1/2}
\label{eq6}
\end{equation}
respectively. Here $\pi_{\rm rel}={\rm AU}(1/D_{\rm L}-1/D_{\rm S})$
denotes the relative lens-source parallax.

The Bayesian analysis of each event was conducted by producing a large number of artificial
lensing events.  For the individual artificial events, the locations and velocities of the 
lenses and source stars were derived from a Galactic model, and the masses of the lenses were 
derived from a model mass function by conducting a Monte Carlo simulation.  In the simulation, 
we adopt the Galactic model and mass function described in detail by \citet{Jung2021}. We then 
construct the posteriors of $M$ and $\dl$ by imposing a weight of $w_i=\exp(-\chi^2/2)$ to each 
simulated event.  Here $\chi^2$ value is computed by
\begin{equation}
\chi^2 = 
{(t_{{\rm E},i} - t_{\rm E})^2 \over [\sigma(\te)]^2} +
{(\theta_{{\rm E},i} - \theta_{\rm E})^2 \over [\sigma(\thetae)]^2},
\label{eq7}
\end{equation}
where $(t_{{\rm E},i}, \theta_{{\rm E},i})$ denote the event time scale and Einstein radius of 
each simulated lensing event computed from the relations in Equation~(\ref{eq6}), $(\te, 
\thetae)$ indicate the measured values, and $[\sigma(\te), \sigma(\thetae)]$ represent their 
measurement uncertainties.  In our analyses, we use the values of $\thetae$ estimated from the 
colors and magnitudes of the primary source stars.

\begin{figure}[t]
\includegraphics[width=\columnwidth]{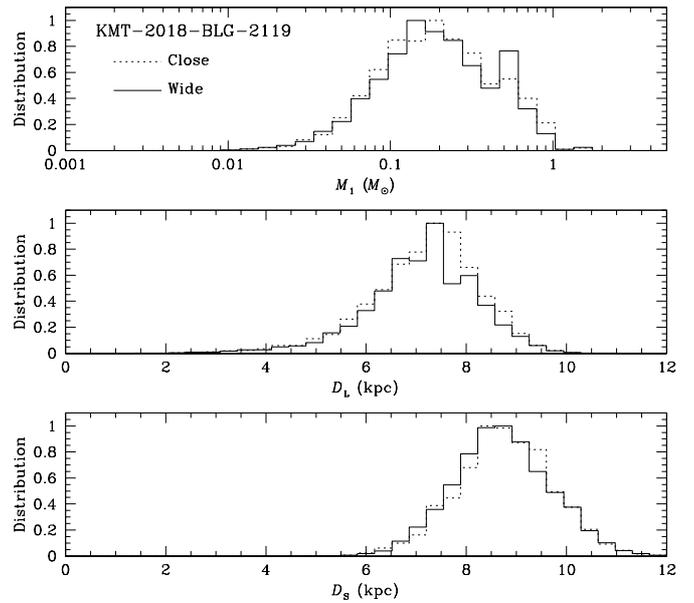}
\caption{
Bayesian posteriors of the physical lens parameters for KMT-2018-BLG-2119.  Notations are 
the same as those in Fig.~\ref{fig:eight}. 
}
\label{fig:nine}
\end{figure}

\begin{table}[t]
\small
\caption{Physical lens parameters\label{table:six}}
\begin{tabular*}{\columnwidth}{@{\extracolsep{\fill}}lllllllr}
\hline\hline
\multicolumn{1}{c}{Parameter}           &
\multicolumn{2}{c}{OGLE-2018-BLG-0584}  &
\multicolumn{2}{c}{KMT-2018-BLG-2119}   \\
\multicolumn{1}{c}{}                    &
\multicolumn{1}{c}{Close}               &
\multicolumn{1}{c}{Wide}                &
\multicolumn{1}{c}{Close}               &
\multicolumn{1}{c}{Wide}                \\
\hline
 $M_1$ ($M_\odot$)  &   $0.22^{+0.33}_{-0.12}$  &  $0.38^{+0.36}_{-0.22}$     &  $0.21^{+0.32}_{-0.12}   $       &  $0.21^{+0.34}_{-0.12}   $    \\    [0.8ex]
 $M_2$ ($M_\odot$)  &   $0.13^{+0.20}_{-0.07}$  &  $0.42^{+0.39}_{-0.24}$     &  $0.012^{+0.018}_{-0.007}$       &  $0.013^{+0.021}_{-0.007}$    \\    [0.8ex]
 $\dl$ (kpc)        &   $6.51^{+1.09}_{-1.44}$  &  $6.75^{+1.03}_{-1.36}$     &  $7.46^{+0.91}_{-1.18}   $       &  $7.38^{0.94}_{-1.09}   $    \\    [0.8ex]
 $a_\perp$ (AU)     &   $1.04^{+0.18}_{-0.23}$  &  $5.24^{+0.80}_{-1.06}$     &  $0.80^{+0.10}_{-0.13}   $       &  $3.09^{+0.40}_{-0.46}   $    \\    [0.8ex]
\hline
\end{tabular*}
\end{table}

The posteriors of the mass of the primary lens and distance to the lens systems constructed 
from the Bayesian analyses for the events OGLE-2018-BLG-0584 and KMT-2018-BLG-2119 are 
presented in Figures~\ref{fig:eight} and \ref{fig:nine}, respectively. In each panel, we draw 
two distributions, in which the dotted and solid ones represent the distributions obtained based 
on the close and wide solutions, respectively. The distribution of the source distance $\ds$ in 
the bottom panel of each figure is presented to show the relative locations of the lens and source.

In Table~\ref{table:six}, we list the Bayesian estimates of the primary and secondary lens masses, 
$M_1$ and $M_2$, distance $\dl$, and projected separation $a_\perp$ between the lens components
for the individual events.  The projected separation is computed from the binary separation, 
angular Einstein radius, and lens distance by $a_\perp =s\dl\thetae$.  The upper and lower limits 
of the individual lens parameters are set as the 16\% and 84\% ranges of the posterior distributions. 
We note that there are some variations in the parameters depending on the close and wide solutions 
of the events. Nevertheless, the estimated parameters indicate that the lens of OGLE-2018-BLG-0584 
is a binary composed of two M dwarfs, and that of KMT-2018-BLG-2119 is a binary composed of a 
low-mass M dwarf and a brown dwarf. The detection of the brown-dwarf companion to KMT-2018-BLG-2119L 
demonstrates the usefulness of binary-lens events in detecting microlensing brown dwarfs as recently 
demonstrated by \citet{Han2022c} and \citet{Han2022e}.

\section{Summary}\label{sec:six}

We reanalyzed the two lensing events OGLE-2018-BLG-0584 and KMT-2018-BLG-2119, for which
there had been no suggested models explaining the anomalies in the lensing light curves. 
It was found that the light curves could not be explained by the usual models based on 
either a 2L1S or a 1L2S interpretation.

We reanalyzed the light curves of the events with more sophisticated models including 
an extra lens or a source component to the 2L1S lens-system configuration. From these 
analyses, we found that a 2L2S interpretation well explained the light curves of both 
events, for each of which there existed a pair of solutions resulting from the 
close--wide degeneracy.

The two events are the sixth and seventh identified events for which both the lens and
source are binaries. For the event OGLE-2018-BLG-0584, the source is a binary composed of two
K-type stars, and the lens is a binary composed of two M dwarfs. For the event KMT-2018-BLG-2119, 
the source is a binary composed of two dwarfs of G and K spectral types, and the lens is a 
binary composed of a low-mass M dwarf and a brown dwarf.

\begin{acknowledgements}
Work by C.H. was supported by the grants of National Research Foundation of Korea 
(2020R1A4A2002885 and 2019R1A2C2085965).
J.C.Y. acknowledges support from U.S. NSF Grant No. AST-2108414.
Y.S. acknowledges support from BSF Grant No. 2020740.
This research has made use of the KMTNet system operated by the Korea Astronomy and Space 
Science Institute (KASI) and the data were obtained at three host sites of CTIO in Chile, 
SAAO in South Africa, and SSO in Australia.
\end{acknowledgements}

\end{document}